\documentclass[onecolumn,notitlepage,nofootinbib]{revtex4-1}

\usepackage{amsmath,amssymb,bbm,soul}
\usepackage{amsthm}
\usepackage{caption}
\usepackage{graphicx}
\usepackage[colorlinks=true,citecolor=blue,linkcolor=blue,urlcolor=blue]{hyperref}

\catcode`\|=\active \def|{
\fontencoding{T1}\selectfont\symbol{124}\fontencoding{\encodingdefault}}

\newcommand{\mathd}{\mathrm{d}}

\newcommand{\tmop}[1]{\ensuremath{\operatorname{#1}}}

\usepackage{xcolor}

\newcommand{\id}{\mathbbm{1}} 

\renewcommand{\BibitemShut}[1]{}
\usepackage[normalem]{ulem}
\newcommand{\stkout}[1]{\ifmmode\text{\sout{\ensuremath{#1}}}\else\sout{#1}\fi}


\begin{document}
\author{Nina Megier}

\email{nina.megier@mi.infn.it}

\affiliation{Dipartimento di Fisica “Aldo Pontremoli”, Università degli Studi di Milano, via Celoria 16, 20133 Milan, Italy}
\affiliation{ Istituto Nazionale di Fisica Nucleare, Sezione di Milano, via Celoria 16, 20133 Milan, Italy}

\author{Andrea Smirne}
\author{Bassano Vacchini}

\affiliation{Dipartimento di Fisica “Aldo Pontremoli”, Università degli Studi di Milano, via Celoria 16, 20133 Milan, Italy}
\affiliation{ Istituto Nazionale di Fisica Nucleare, Sezione di Milano, via Celoria 16, 20133 Milan, Italy}

\begin{abstract}
Using a newly introduced connection between the local and non-local description of open quantum system dynamics, we investigate the relationship between these two characterisations in the case of quantum semi-Markov processes. This class of quantum evolutions, which is a direct generalisation of the corresponding classical concept, 
guarantees mathematically well-defined master equations,
while accounting for a wide range of phenomena, possibly in the non-Markovian regime. In particular, we analyse the emergence of
a dephasing term when moving from one type of master equation to the other, by means of several examples. We also investigate the corresponding Redfield-like approximated dynamics, which are obtained after a coarse graining in time. Relying on general properties of the associated classical random process, we conclude that such an approximation always leads to a Markovian evolution for the considered
class of dynamics.

\end{abstract}

\title{Evolution equations for quantum semi-Markov dynamics}

\maketitle

\section{Introduction}\label{sec:int}
The coupling of a quantum system to some external degrees of freedom can rarely be neglected. Since many valuable quantum features, such as non-classical correlations and superpositions, are very fragile and sensitive to such perturbation, understanding the dynamics of open quantum systems is highly relevant not only from a fundamental point of view but also for implementations of quantum technologies \cite{Arndt2011a, Acin2018,Wang2020a}. Especially going beyond the well understood regime of evolutions characterised by the semigroup property, that is in Gorini-Kossakowski-Sudarshan-Lindblad (GKSL) form \cite{Gorini1976, lindblad1976}, has recently attracted great attention. Non-Markovian effects corresponding to such  a type of dynamics were shown to be of advantage e.g. for quantum control tasks \cite{reich2015, PhysRevLett.111.010402} and quantum information \cite{bylicka2014,Cialdi2017a}.

The dynamics of open quantum systems can be described in terms of local and non-local master equations, i.e. evolution equations for the reduced density operator $\rho_t$ \cite{breuerbook}. Both characterisations are in principle equivalent, as they contain the  whole information about the reduced dynamics, nonetheless the knowledge of both can be of advantage. This is the case since some pieces of information are more easily obtained from one rather than the other description. As a relevant example, a particular divisibility property of the dynamical map $\Lambda_t$ determining the evolved state according to $\Lambda_t[\rho_0]=\rho_t$ can be directly concluded from the Lindblad operator form of the local description \cite{hallcresserandersson}; on the other hand the physical origin of the evolution equation can be sometimes better visible from the non-local characterisation. This is the case for the quantum semi-Markov evolutions \cite{Budini2004,breuer2008,PhysRevE.79.041147,bassano_2012,PhysRevA.94.020103, PhysRevLett.117.230401}, which we will investigate in this paper, as they provide one of the few known classes of quantum non-Markovian evolutions which have been thoroughly studied and are rich enough to put into evidence the interplay between local and non-local description. 

The semi-Markov dynamics is characterised by two main ingredients: quantum  evolutions continuous in time and random quantum jump processes. The jumps interrupt the continuous evolution at random times. The definition of quantum semi-Markov process is a direct generalisation of the classical concept. However, the transition to the quantum realm makes the theory reacher and more challenging, as here the operator ordering plays an important role. The quantum semi-Markov dynamics can be highly non-Markovian \cite{Vacchini2011}, but their structure guarantees that they are indeed proper quantum evolutions, i.e. complete positive (CP) and trace preserving (T). What is more, they describe many relevant physical systems and are used in many applications, as micromaser \cite{PhysRevA.46.5913,PhysRevA.52.602,Cresser2019a}, quantum thermometry \cite{PhysRevLett.88.097905, PhysRevLett.123.180602} or general collision models \cite{Ciccarello2013a,Lorenzo2017a}, making them an important playground to investigate non-Markovian effects.  

In this paper we study the connection and the interplay between the local and non-local master equations for quantum semi-Markov dynamics. In Section \ref{sec:TCLvsNZOld} we summarise the known connections between these two characterisations of quantum evolutions for general systems.
The class of quantum semi-Markov dynamics to be considered is introduced in Section \ref{sec:random}. In Section \ref{sec:Lindblad} the
interplay between local and non-local generators for these dynamics is discussed by means of example, further putting into evidence the role of expressions in Lindblad operator form. In Section \ref{sec:interplay} we discuss a generally valid approximation of the considered class of evolutions, which takes the form of a Redfield-like approximation and always leads to a well-defined quantum dynamics due to general properties of classical waiting time distributions. Eventually, in Section \ref{sec:Concl}, we summarise our findings.

\section{Local and non-local representations of open quantum system dynamics}\label{sec:TCLvsNZOld}

When investigating an open quantum system, i.e. a quantum system interacting with some external degrees of freedom, one is mostly interested in the reduced observables associated
with the open system only. Their statistics is fixed by the reduced density operator $\rho_t$, making it one of the central objects in the theory of open quantum systems \cite{breuerbook}. The corresponding evolution equations can have two forms: local, also called time-convolutionless (TCL) \cite{Hashitsumae1977a,Shibata1978a},
\begin{eqnarray}\label{eq:tcl}
\frac{\mathd}{\mathd t} \rho_t = \mathcal{K}^{\tmop{TCL}}_{t} \rho_t,
\end{eqnarray}
and non-local, also called Nakajima-Zwanzig (NZ)  \cite{nakajima1958,zwanzig1960}
\begin{eqnarray}\label{eq:nz}
\frac{\mathd}{\mathd t} \rho_t = \int_0^t \mathd \tau \mathcal{K}^{\tmop{NZ}}_{t-\tau} \rho_\tau\equiv (\mathcal{K}^{\tmop{NZ}} * \rho)_t.
\end{eqnarray}
Both equations are equivalent, in the sense that their solutions give the same object $\rho_t$, however, their structure is significantly different. A well-known example of the first type is the GKSL master equation, where the generator $\mathcal{K}^{\tmop{TCL}}_{t}$ is time-independent \cite{Gorini1976, lindblad1976}. One can obtain such a 
form of the evolution from the microscopic model that fixes the environment and system-environment
interaction by conducting the Born-Markov approximation \cite{breuerbook}, which is based on the separation of relevant time scales of the system and its environment. The time-dependent generalisations of the GKSL equation, such that the same structure is preserved, are often introduced on phenomenological grounds, which is a strategy with several pitfalls \cite{PhysRevA.89.042117, bylicka}. On the other hand, both local and non-local equations can be obtained for general open quantum systems in terms of projection operator techniques. However, the exact calculation of the quantities occurring in the resulting expressions is in general not possible. That is why in practice one mostly has to resort to perturbative techniques \cite{Haake1973,breuerbook}.

One could wonder if the non-local equation has more general validity, while the local equation can only be written in some limited cases: the basic issue here is invertibility of the evolution map, generally granted only up to a given time \cite{Breuer2001b,Vacchini2010b}. However, under this mild condition the equation of the form \eqref{eq:tcl} can always be obtained from the non-local description, as
\begin{eqnarray}
  \label{eq:start}
\frac{\mathd}{\mathd t} \rho_t = \int_0^t \mathd \tau \mathcal{K}^{\tmop{NZ}}_{t-\tau} \rho_\tau= \int_0^t \mathd \tau \mathcal{K}^{\tmop{NZ}}_{t-\tau} \Lambda_\tau \Lambda^{-1}_t\rho_t= \mathcal{K}^{\tmop{TCL}}_{t} \rho_t.
\end{eqnarray}
By noting that the dynamical map $\Lambda_t$ satisfies the same evolution equations as the reduced density operator $\rho_t$ ($\dot{\rho}_t=\dot{\Lambda}_t[\rho_0]$), this leads us to the connection between the local generator $\mathcal{K}^{\tmop{TCL}}_t$ and the dynamical map,
\begin{eqnarray}
\mathcal{K}^{\tmop{TCL}}_t =
  \dot{\Lambda}_t \Lambda^{-1}_t \nonumber,
\end{eqnarray}
where, again, the invertibility of the dynamical map $\Lambda_t$ was assumed. When the invertibility condition is not satisfied, the local description can nonetheless exist, under certain consistency conditions \cite{Andersson2007}.

The non-local generator can also be expressed directly in terms of the dynamical map.
In the time domain the relationship involves the first and second derivative of the evolution map
\begin{eqnarray}
  \label{eq:NZconnLambda1}
  \mathcal{K}^{\tmop{NZ}}_{t} =
  \ddot{\Lambda}_t - ( \dot{\Lambda} \ast \mathcal{K}^{\tmop{NZ}} )_t=\ddot{\Lambda}_t - ( \mathcal{K}^{\tmop{NZ}} \ast \dot{\Lambda})_t,
\end{eqnarray}
while in Laplace transform we have the identities expressed in terms of the transforms of the evolution map or of its first derivative \begin{eqnarray}
  \widetilde{\mathcal{K}^{\tmop{NZ}}}_u = \frac{u \tilde{\Lambda}_u -
  \mathbbm{1}}{\tilde{\Lambda}_u}
  = \frac{u \tilde{\dot{\Lambda}}_u}{\mathbbm{1} +
  \tilde{\dot{\Lambda}}_u}
 , \label{eq:NZconnLambda} 
\end{eqnarray}
where $\widetilde{\Phi}_u$ denotes the Laplace transform of the operator $\Phi_t$.
The last equality in \eqref{eq:NZconnLambda1} is a consequence of the initial condition $\Lambda_0=\id$. As a result, in some expressions containing the dynamical map one can act as if the operator ordering does not matter. This was used for example in \cite{Kidon2018}, where the advantage of Eq.~\eqref{eq:NZconnLambda1} for numerical calculations of the memory kernel was shown in an example related to electron transport. 

From the previous relations, a direct connection between the local
and the non-local generators can be derived, as
\begin{eqnarray}\label{eq:NZvsTCLMap}
 \widetilde{\mathcal{K}^{\tmop{NZ}}}_u = \frac{u \widetilde{(\mathcal{K}^{\tmop{TCL}} \Lambda)}_u}{\mathbbm{1} + \widetilde{(\mathcal{K}^{\tmop{TCL}} \Lambda)}_u}, 
\end{eqnarray}
i.e. to get the non-local generator one has to know the product of the local generator and the dynamical map. This is of course not optimal, as the knowledge of the dynamical map is needed. In the situations where the solution of Eqs.~\eqref{eq:tcl} and \eqref{eq:nz} is already known, the usefulness of Eq.~\eqref{eq:NZvsTCLMap} is, however, rather limited.

Recently, a different relation was introduced, providing a direct connection between non-local and local generator, namely  starting from Eq.~\eqref{eq:start} and using the following representation of the dynamical map in terms of the local generator
\begin{eqnarray}
  \label{eq:3}
  \Lambda_t=\mathcal T_{\leftarrow}  e^{\int\limits_0^t d\tau \mathcal{K}^{\tmop{TCL}}_{\tau}}
\end{eqnarray}
one obtains the expression \cite{nestmann2020quantum}
\begin{eqnarray}\label{eq:TCLvsNZTimeOrdering}
\mathcal{K}^{\tmop{TCL}}_t = \int\limits_0^t ds  \mathcal{K}^{\tmop{NZ}}_{t-s} \mathcal T_{\rightarrow}  e^{-\int\limits_s^t d\tau \mathcal{K}^{\tmop{TCL}}_{\tau}}, 
\end{eqnarray}
where $\mathcal T_{\leftarrow}$ ($\mathcal T_{\rightarrow}$) denotes (inverse) time ordering.
Note that the connection between local and non-local generators given by Eq.~\eqref{eq:TCLvsNZTimeOrdering}, while being implicit, can be understood as a fixed-point relation. Though mathematically involved, it already proved advantageous for numerical calculations.

A powerful connection between the local and non-local generators can be obtained with the damping-basis representation \cite{Briegel1993}, when one restricts to (diagonalisable) commutative dynamics, i.e. satisfying 
\begin{eqnarray}
  \label{comm}
[\Lambda_t,\Lambda_s]=0
\end{eqnarray}
with $[\cdot,\cdot]$ being the commutator \cite{Chruscinski2010,Chruscinski2014}.  In \cite{megier2020interplay} it was shown, that in this case the local and non-local generators can be written as
    \begin{eqnarray} \label{eq:dampbasTCLNZ}
\mathcal{K}^{\tmop{TCL}}_t = \sum_{\alpha}^{} m^{\tmop{TCL}}_{\alpha}(t) \mathcal{M}_{\alpha}, && \mathcal{K}^{\tmop{NZ}}_t = \sum_{\alpha}^{} m^{\tmop{NZ}}_{\alpha}(t) \mathcal{M}_{\alpha},
\end{eqnarray}
where $m^{\tmop{TCL}}_{\alpha}(t)$ and $m^{\tmop{NZ}}_{\alpha}(t)$
are functions of time (the eigenvalues of the corresponding damping-basis decompositions) and
are related by
\begin{eqnarray}
  m^{\tmop{NZ}}_{\alpha}(t) &=& \mathfrak{I}\left(\frac{u \widetilde{G_{\alpha}}(u)}{1+\widetilde{G_{\alpha}}(u)}\right)(t),
                                \label{eq:mnz}
                                \\
                                m^{\tmop{TCL}}_{\alpha}(t) &=& \frac{G_{\alpha} (t)}{1 + \int_0^t \mathd \tau G_{\alpha} (\tau)},
                                                               \label{eq:tcltcl}\\
  \text{with \hspace{0.5cm}}
G_{\alpha}(t) &\equiv& \frac{\mathd}{\mathd t} e^{\int_0^t \mathd \tau m^{\tmop{TCL}}_{\alpha}(\tau)}= \mathfrak{I}\left(\frac{\widetilde{m^{\tmop{NZ}}_{\alpha}}(u)}{u-\widetilde{m^{\tmop{NZ}}_{\alpha}}(u)}\right)(t),  \label{eq:mtcl}
\end{eqnarray}
where $\widetilde{f}(u)$ denotes the Laplace transform of the function $f(t)$, while $\mathfrak{I} (\widetilde{f}(u))(t)$ denotes the inverse Laplace transform.
What is more, the maps $\mathcal{M}_{\alpha}$ in Eq.~\eqref{eq:dampbasTCLNZ} can be written with bi-orthogonal\footnote{Bi-orthogonal means here, that $\langle \varsigma_{\alpha} , \tau_{\beta} \rangle = \langle \tau_{\alpha} , \varsigma_{\beta} \rangle = \delta_{\alpha \beta}$ is satisfied, where we consider the standard scalar product in the Hilbert-Schmidt space of linear operators defined as $\langle \omega , \sigma \rangle = \text{Tr} \,\omega^\dag \sigma$.
} bases $\left\{\tau_{\alpha}\right\}$
and $\left\{\varsigma_{\alpha}\right\}$
of operators acting on the open-system Hilbert space (the damping bases of the generators),
 as
\begin{eqnarray*}
\mathcal{M}_{\alpha}[\cdot]= \text{Tr}[\varsigma^\dagger_{\alpha} \cdot]\tau_{\alpha},
\end{eqnarray*}
and, because of the commutativity of the dynamics, they are time-independent. Accordingly, the operational form of the local and the non-local generators is the same in this representation, and the direct connection between the time-dependent functional terms is given. Though the inverse Laplace transform in Eq.~\eqref{eq:mnz} in general cannot be calculated, the above link between the two characterisations is not only a formal one. In \cite{megier2020interplay} it was shown that it enables to understand the relations between the Lindblad operator form of the two generators,
as well as the connection between the (non-)Markovianity of the original and the Redfield-like approximated dynamics.

Indeed one of the major motivations for addressing both local and non-local formulations of the dynamics is the fact that they both allow describing dynamics beyond the semigroup paradigm, but provide different insights with respect to the different approaches to non-Markovianity in the quantum regime.
In general the classical definition of non-Markovianity cannot be straight-forwardly transferred to the quantum regime. That is why many non-equivalent definitions of quantum Markovianity exist, see e.g. the reviews \cite{Rivas_2014, Breuer2016a,Devega2017a,LiHallWiseman2017}. In the present contribution we are mainly concerned with the operator structure of models which have a common root in a classical description, that of semi-Markov processes.  For a particular subset of semi-Markov processes following Eq. \eqref{eq:nzSemiMark}, we will show that the Redfield-like approximated dynamics is Markovian since the associated dynamical map $\Lambda_t$  has the property that the transformation $\Lambda_{t,s}$ satisfying $\Lambda_t=\Lambda_{t,s}\Lambda_{s}$ is a CP map\footnote{Note that in this case a CP-divisibility is equivalent to P-divisibility.} for $0\leq s \leq t$. We stress that while CP-divisibility already implies lack of information backflow, P-divisibility appears to play in general a distinguished role. Indeed, P-divisibility, besides a definite mathematical characterisation, has a clear physical meaning: On the one hand, it can be brought in connection with the information backflow from the environment into the reduced system, becoming manifest by a non-monotonic behaviour of the trace distance between two quantum states of the system \cite{Bassano2015}; on the other hand, it allows one to interpret the dynamics as the result of a continuous measurement performed on the open system \cite{Smirne2020}.

The Redfield-like approximation can be obtained from the non-local description \eqref{eq:nz} by the following coarse graining in time
\begin{eqnarray}\label{eq:red}
\frac{\mathd}{\mathd t} \Lambda^{\tmop{Red}}_t 
= \mathcal{K}^{\tmop{Red}}_t \Lambda^{\tmop{Red}}_t,
\end{eqnarray}
with
\begin{eqnarray}
  \label{eq:7}
 \mathcal{K}^{\tmop{Red}}_t
= \int_0^t \mathd \tau \mathcal{K}^{\tmop{NZ}}_{\tau},
\end{eqnarray}
which accordingly provides us with an approximated local equation, starting from the exact non-local one.
A key point is that the Redfield-like approximated evolution also shows the same structure as the exact local and non-local equations:
\begin{eqnarray}
\mathcal{K}^{\tmop{Red}}_t &=& \sum_{\alpha}^{} m^{\tmop{Red}}_{\alpha}(t) \mathcal{M}_{\alpha}, \label{eq:kred}
\end{eqnarray}
with
\begin{eqnarray}
m^{\tmop{Red}}_{\alpha}(t) &=& \int_0^t \mathd \tau m^{\tmop{NZ}}_{\alpha}(\tau) \label{eq:mred},
\end{eqnarray}
which simplifies the analysis of the connection between the (non-)Markovianity of the original and the Redfield-like approximated dynamics.

\section{Quantum semi-Markov evolutions}\label{sec:random}

In this paper, we will focus our analysis on a class of quantum semi-Markov evolutions.
While a strictly unique definition of quantum semi-Markov evolution is missing, this term is used for the quantum counterpart of classical semi-Markov processes \cite{Feller1968,Ross2003}, which arises merging renewal processes and Markovian jump processes. Thus, a classical semi-Markov process describes a random evolution characterised by transitions between a fixed set of states according to possibly site-dependent waiting time distributions.
The latter describes the random times spent in a site before jumping, with some fixed transition probabilities, to one of the other sites. Moving to the quantum framework, transitions are replaced by quantum jumps described by CPT maps, and the possible evolution in between the jumps is described by a time-dependent collection of CPT maps. The general form of such evolutions can thus be written as
\begin{eqnarray}\label{eq:semiMarkGen}
\rho_t= p_0(t) \mathcal{F}(t) \rho_t+  \sum\limits_{n=0}^\infty \int\limits_0^t dt_n... \int\limits_0^{t_2} dt_1  p_n(t; t_n,...,t_1) \mathcal{F}(t-t_n)\mathcal{E} ...\mathcal{E} \mathcal{F}(t_1) \rho_0,
\end{eqnarray}
where $\mathcal{E}$ is a jump operator (a CPT map),  $\{\mathcal{F}(t)\}_{t \geq 0}$ is a family of CPT maps describing the evolution
between the jumps and $p_n(t; t_n,...,t_1)$ are the probability densities for $n$ jumps at fixed times $t_n,...,t_1$.
In order to comply with classical semi-Markov processes, these probability densities correspond to a renewal process and can therefore be expressed in the form
\begin{eqnarray*}
p_n(t; t_n,...,t_1)=f(t-t_n)... f(t_2-t_1)g(t_1),
\end{eqnarray*}
where $f(t)$ is a waiting time distribution, that is a probability density over positive times, and $g(t)$ is the corresponding survival probability given by 

\begin{eqnarray}\label{eq:extra1}
g(t)=1-\int\limits_0^t f(s)ds.
\end{eqnarray}
For simplicity we have assumed that $\mathcal{E}$, $\mathcal{F}(t)$ and the waiting time distributions $f(t)$ are the same at each step,
even though a more general description can be considered \cite{breuer2008,Vacchini2013a,PhysRevLett.117.230401}. In particular, the intermediate time evolution can be fixed to be of the exponential type, thus focusing on the
role of different waiting times and jump operators, and giving rise to a class of non-Markovian dynamics which has been termed quantum renewal processes \cite{PhysRevLett.117.230401,Vacchini2020a}. A
particular subset of quantum semi-Markov processes can be obtained by assuming 
$\mathcal{F}(t) \rightarrow \mathbbm{1}$, so that
\begin{eqnarray}\label{eq:semiMarkSimp}
  \rho_t= \Lambda_t[\rho_0]=\sum\limits_{n=0}^\infty  p_n(t) \mathcal{E}^n\rho_0        
\end{eqnarray}
where $p_n(t)$ is the probability density for having $n$ jumps in a time $t$. The corresponding non-local evolution equation \eqref{eq:nz} reads in this case \cite{Budini2004,Vacchini2011}
\begin{eqnarray}\label{eq:nzSemiMark}
\frac{d}{dt}\rho_t= \int\limits_0^t ds\, k(t-s) ( \mathcal{E}-\mathbbm{1})\rho_s,
\end{eqnarray}
where $k(t)$ is a memory kernel uniquely determined by the waiting time distribution $f(t)$ according to
\begin{eqnarray}
  \label{eq:1}
  f(t)=\int\limits_0^t ds\, k(t-s)g(s).
\end{eqnarray}
The operatorial form of the memory kernel $\mathcal{K}^{\tmop{NZ}}_t$ is then determined by the jump operator $\mathcal{E}$ and the functional time-dependence by the  waiting time distribution $f (t)$. The dynamics is accordingly commutative in the sense of Eq. \eqref{comm}.

As the jump map $\mathcal{E}$ is CPT, it can be written in Kraus form
\begin{eqnarray}
  \label{eq:kraus}
 \mathcal{E}(\omega)=\sum\limits_i C_i \omega C_i^\dagger ,
\end{eqnarray}
with $\sum\limits_i  C_i^\dagger C_i=\id$. Then, it is immediately 
evident that the operator appearing in Eq.~\eqref{eq:nzSemiMark} has the GKSL form
 \begin{eqnarray}
   \label{eq:5}
   \mathcal{L}=\mathcal{E}-\mathbbm{1},
 \end{eqnarray}
where the Kraus operators $C_i$ play the role of Lindblad operators.  
As said before, quantum semi-Markov processes provide a generalisation of the classical concept \cite{Feller1968,Ross2003}. Nonetheless, the quantum class is more complex,
due to the non-trivial role played by the operator ordering  \cite{PhysRevLett.117.230401,PhysRevA.94.020103,Vacchini2020a};
quantum semi-Markov processes provide a further instance of how the notion of 
Markovianity cannot be naively transferred from the realm of classical 
stochastic processes to the one of open quantum system dynamics \cite{Vacchini2011}.\\

\section{Lindblad operator form}\label{sec:Lindblad}

For a given dynamics different evolution equations can be considered, both local and non-local according to \eqref{eq:tcl} and \eqref{eq:nz}, respectively. Moreover in both cases a gauge freedom is available, so that the operator structure is not uniquely fixed. 
However, starting from the damping basis decomposition given by Eq.~\eqref{eq:dampbasTCLNZ} one can bring both generators in Lindblad operator form, as shown in 
\cite{megier2020interplay}. With the Lindblad operator form of the evolution equation we mean the one directly generalising the well-known GKSL master equation, i.e.
\begin{eqnarray}\label{eq:gorinidiag}
\mathcal{K}_t (\rho) = -i \left[ H(t) , \rho \right] + \sum^{}_{\alpha} \gamma_{\alpha}(t) \mathcal{L}_{L_\alpha(t)}(\rho)  ,
\end{eqnarray}
with
\begin{eqnarray}\label{eq:gorinidiag2}
\mathcal{L}_{X}(\omega) = X \omega X^{\dag}
-\frac{1}{2} \left\{X^{\dag} X , \omega \right\},
\end{eqnarray}   
where the Lindblad operators $L_{\alpha}(t) $ and damping rates $\gamma_{\alpha}(t) $ are time-dependent. The damping rates  $\gamma_{\alpha}(t) $ can be negative, which makes the corresponding dynamical map non CP-divisible \cite{hallcresserandersson}. This clear connection to this definition of non-Markovianity is one of the reasons for which this form of the master equation is widely used.

Importantly, while the damping-basis representation \eqref{eq:dampbasTCLNZ} provides
the local and non-local generators with the same operatorial structure, 
this is generally not the case for the Lindblad operator form. 
A situation where also the Lindblad operator form is the same for the two generators
is when it has
one single overall time-dependent rate and only one (possibly degenerate) non-zero eigenvalue in the damping-basis representation, that is
\begin{eqnarray}
  \label{eq:2}
  \mathcal{K}^{\text{TCL}}_t  = \gamma (t) \mathcal{L}
\end{eqnarray}
with $\mathcal{L}=\ell \sum\limits_\alpha\mathcal{M}_\alpha$, so that
\begin{eqnarray}
  \label{eq:6}
  \mathcal{K}^{\text{NZ}}_t  = \frac{m^{\text{NZ}}(t)}{\ell}\mathcal{L}.
\end{eqnarray}
As commonly one (or both) of these conditions is (are) violated, in general some Lindblad operators contained in the local description can be missing in the non-local one, and vice versa \cite{megier2020interplay}. This makes the interpretation of the underlying physical origin of the dynamics more difficult. Here, we want to present two examples of such a phenomenon making reference to the class of quantum semi-Markov evolutions.

\subsection {Addition of dephasing in the local generator}\label{sec:nzTOtcl}

In the first example we start with the following non-local generator 
\begin{eqnarray} \label{eq:exnz1} 
\mathcal{K}^{\tmop{NZ}}_t (\omega) = k (t)  \left( \sigma_-
  \omega \sigma_+ + \sigma_+ \omega \sigma_- - \omega \right),
\end{eqnarray}
which is of the form given by Eq.~\eqref{eq:nzSemiMark}, where the Kraus operators corresponding to the jump operator $\mathcal{E}$, see Eq.~\eqref{eq:kraus}, are given by $C_1=\sigma_-$ and $C_2=\sigma_+$. It is important to stress that this kernel indeed provides a well-defined semi-Markov dynamics whenever $k(t)$ can be interpreted as  classical memory kernel determined by a waiting time distribution as in Eq.~\eqref{eq:1} \cite{Budini2004}. 
The non-local generator in Eq.~\eqref{eq:exnz1} can be written as the sum of two generators in Lindblad operator form describing the gain or loss of an excitation by a qubit, with the same time-dependent prefactor $k(t)$.
To obtain the corresponding local generator, we use the results obtained via the damping bases in Section \ref{sec:TCLvsNZOld}. The damping bases $\left\{\tau_\alpha\right\}_{\alpha=1,\ldots, 4} $ and $\left\{\varsigma_\alpha\right\}_{\alpha=1,\ldots, 4} $ coincide in this instance, as the generator $\mathcal{K}^{\tmop{NZ}}_t (\omega) $ is self-adjoint \cite{megier2020interplay}, and they read $(1/\sqrt{2})(\mathbbm{1},\sigma_x,\sigma_y,\sigma_z)$. 
Moreover, Eqs.~\eqref{eq:mnz}  and \eqref{eq:tcltcl}, which relate the eigenvalues of the non-local and local generator via \eqref{eq:mtcl}, lead for the present case to the expressions
\begin{eqnarray}
\left\{m^{\tmop{NZ}}_\alpha(t)\right\}_{\alpha=1,\ldots, 4} &=& \left\{0,-k(t),-k(t),-2 k(t)\right\},\label{eq:evalNZ1}\\
\left\{m^{\tmop{TCL}}_\alpha(t)\right\}_{\alpha=1,\ldots, 4} &=& \left\{0,-\frac{1}{2}(h(t)+\mu(t)),-\frac{1}{2}(h(t)+\mu(t)),-2 \mu(t)\right\}\label{eq:evalTCL1}.
\end{eqnarray}
Note that both generators have two different non-zero eigenvalues; the non-local generator can be written as $\mathcal{K}^{\tmop{NZ}}_t=k(t)\mathcal{L}$, with $\mathcal{L}$ a generator in  GKSL form, but the relation between Eqs.~\eqref{eq:2} and Eq.~\eqref{eq:6} does not apply. 

Most importantly, each of the quantities defining the eigenvalues in Eqs.~(\ref{eq:evalNZ1}) and (\ref{eq:evalTCL1}) allows for a natural probabilistic interpretation.  As said before, $k(t)$ in Eq.~(\ref{eq:evalNZ1}) is uniquely determined by the waiting time distribution $f(t)$ via Eq. (\ref{eq:1}). In addition, the quantities $h(t)$ and $\mu(t)$, occurring in Eq.~\eqref{eq:evalTCL1}, have the following meaning.  The first one, $h(t)$, is the hazard rate, a positive function given by the ratio between the waiting time distribution $f(t)$ and its associated survival probability $g(t)$, see Eq.~(\ref{eq:extra1}),
\begin{eqnarray}
  \label{eq:h}
  h(t)=-\frac{\dot{g}(t)}{g(t)},
\end{eqnarray}
where we have used the fact that $f(t)=-\dot{g}(t)$;
in other terms, the hazard rate is proportional to the logarithmic derivative of the survival probability.
Moreover, note that the hazard rate is not a probability density, since it is not normalised; nonetheless, it can be interpreted  as a measure of jump probability: the greater the hazard rate in some time interval, the greater the probability of jump in this time interval \cite{Ross2003}. On the other hand, the function $\mu(t)$ reads 
\begin{eqnarray}\label{eq:gamma}
  \mu(t)=-\frac{1}{2}\frac{\dot{q}(t)}{q(t)},
\end{eqnarray}
so that it is proportional to the logarithmic derivative of the modulus of $q(t)$, which stands for the difference between the probability to have an even or an odd number of jumps at time $t$: $q(t)=\sum\limits_{n=0}^{\infty}p_{2n}(t)-\sum\limits_{n=0}^{\infty}p_{2n+1}(t)$ \cite{bassano_2012}. In contrast to the hazard function $h(t)$ the quantity $\mu(t)$ can turn negative or even diverge  \cite{Vacchini2011}.

Thanks to the relation given by Eq.~\eqref{eq:dampbasTCLNZ}, one can now obtain the local generator, which in the Lindblad operator form reads
\begin{eqnarray} \label{eq:extcl1} 
\mathcal{K}^{\tmop{TCL}}_t (\omega) = \mu (t)  \left( \sigma_-
  \omega \sigma_+ + \sigma_+ \omega \sigma_- - \omega\right)
  + \frac{1}{2}(h(t)-\mu(t))(\sigma_z
  \omega \sigma_z - \omega).
\end{eqnarray}
Accordingly, an additional dephasing term $\mathcal{L}_{\sigma_z}(\omega)$ occurs in the local case, at variance with the non-local one. Note that this contribution only vanishes when the underlying waiting time distribution is of exponential type, that is (classically) Markovian, since only in this case one has $h(t)=\mu(t)$. This change of operator structure from local to non-local representation is therefore a feature associated to classical non-Markovianity, i.e., to the presence of memory in the waiting time distribution.

\subsection {Addition of dephasing in the non-local generator}\label{sec:tclTOnz}

In the next example we encounter the complementary situation in which the non-local generator contains more terms than the local one. 
In this respect, we are leaving the strict framework of semi-Markov processes defined via Eqs.~\eqref{eq:nzSemiMark}, allowing for an additional operator contribution, though the overall derivation is still based on rates obtained from quantities determined by a classical waiting time distribution. To this aim we consider a local generator of the following form

 \begin{eqnarray}
  \label{eq:tcl0} \mathcal{K}^{\tmop{TCL}}_t (\omega) = h(t)  \left( \sigma_- \omega \sigma_+ - \frac{1}{2}  \{ \sigma_+
  \sigma_-, \omega \} \right),
\end{eqnarray}
where $h(t)$ is again a hazard rate, fixed by a waiting time  $f(t)$. Note that $h(t)$ is by construction a positive quantity providing information for a jump to take place in the
subsequent time interval.  An evolution equation of this form appears e.g. when considering a qubit coupled to a bosonic bath at zero temperature \cite{Garraway1997}; finite-temperature baths can be treated via a proper transformation of
the system-environment couplings \cite{Tamascelli2019}. The associated damping bases are given by
\begin{eqnarray}
  \left\{\tau_\alpha\right\}_{\alpha=1,\ldots, 4} = \left\{\frac{\mathbbm{1}-\sigma_z}{2},\sigma_z, \sigma_+, \sigma_-\right\}
  \qquad
\left\{\varsigma_\alpha\right\}_{\alpha=1,\ldots, 4} = \left\{\mathbbm{1},\frac{\mathbbm{1}+\sigma_z}{2},\sigma_+, \sigma_-\right\}, \label{eq:vecdex2}
\end{eqnarray}
with the corresponding eigenvalues for local and  non-local generator, respectively,
\begin{eqnarray}
\left\{m^{\tmop{TCL}}_\alpha(t)\right\}_{\alpha=1,\ldots, 4} &=& \left\{0,-h(t),-\frac{1}{2}h(t),-\frac{1}{2}h(t)\right\}\label{eq:TCLvalueex2}\\
  \left\{m^{\tmop{NZ}}_\alpha(t)\right\}_{\alpha=1,\ldots, 4} &=& \left\{0,-k(t),-k_\surd(t),-k_\surd(t)\right\}.
\label{eq:NZvalueex2}
\end{eqnarray}
The function $k(t)$ is the memory kernel associated to the original waiting time $f(t)$ fixed by $h(t)$. The
function $k_\surd (t)$ also is a memory kernel, uniquely determined by the other one as follows. Notice first that if $g (t)$ is a survival
probability, then also $\sqrt{g (t)}$ is,  since it is still a monotonously decreasing function starting from one. This new survival probability is uniquely associated
with another memory kernel, which we denoted as $k_{\surd} (t)$, so that we have the identities
\begin{eqnarray}
  f(t)=\int\limits_0^t ds\, k(t-s)g(s)\\
  f_\surd(t)=\int\limits_0^t ds\, k_\surd(t-s) \sqrt{g (s)}.
\end{eqnarray}
In this case the operator Lindblad form of the non-local generator contains an additional dephasing channel $\mathcal{L}_{\sigma_z}(\omega)$
\begin{eqnarray}
  \label{eq:k1k2}
\mathcal{K}^{\tmop{NZ}}_t (\omega)=  k(t) \left( \sigma_-  \omega\sigma_+ - \frac{1}{2}  \{
  \sigma_+ \sigma_-, \omega\} \right) + \left(k_\surd(t)-\frac{k(t)}{2}\right)(\sigma_z
  \omega \sigma_z - \omega) . 
\end{eqnarray}
Accordingly, the physical meaning of the particular terms occurring in the Lindblad operator form of local and non-local descriptions is not fixed, since the particular
Lindblad operators are, in general, not preserved by going from one characterisation to the other. Also in this case one can verify that the condition  $k_\surd(t)={k(t)}/{2}$,
guarantying the disappearance of the additional term, is only verified for a Markovian waiting time distribution of exponential form. When considering the solution of the dynamics described by Eq.~\eqref{eq:tcl0}, or equivalently by Eq. \eqref{eq:k1k2}, one can see that the two survival probabilities are related to the decay factors of populations and coherences, which are one the square of the other, as typically happens in amplitude damping channels.

\subsection {Operationally-invariant description}\label{sec:nochange}

The change in operatorial structure when moving from a local to a non-local description expressed in Lindblad operator form, and therefore apparently not amenable to a direct physical interpretation, is obviously not the rule. 
One can also consider situations in which the operator structure remains the same, and only the relation between classical memory kernel and local rates has to be worked out. As to be discussed in the next Section this calls for a complicated and rich relationship between classical functional contribution and operator structure.

We now consider two examples, 
also taken
from the class of semi-Markov processes as in Eq.~\eqref{eq:nzSemiMark}, in which the Lindblad operator structure remains the same in both local and non-local description. 
Consider the two semi-Markov processes
\begin{eqnarray}
\mathcal{E}_{\tmop{diag}}= \sigma_+ \sigma_- \cdot \sigma_+ \sigma_- + \sigma_- \sigma_+ \cdot \sigma_- \sigma_+,\label{eq:diag}
\end{eqnarray}
which performs a diagonalization in the eigenbasis of $\sigma_z$ and corresponds to the choice $C_1=\sigma_+ \sigma_-$ and $C_2=\sigma_- \sigma_+$ in Eq.~\eqref{eq:kraus}, and
\begin{eqnarray} \mathcal{E}_{\tmop{deph}}= \sigma_z \cdot \sigma_z,\label{eq:deph}
\end{eqnarray}
which performs the dephasing operation and is determined by a single operator $C=\sigma_z$.  For both jump maps we consider evolutions with the same classical memory kernel $k(t)$. Using the formalism of Section \ref{sec:TCLvsNZOld}, the local and non-local master equations corresponding to these evolutions are then given by

\begin{eqnarray}\label{eq:hdeph}
  \frac{\mathd}{\tmop{dt}} \rho_t = h(t) [
  \mathcal{E}_{\tmop{diag}} - \mathbbm{1} ] \rho_t
  = \int\limits_0^t ds\, k(t-s) ( \mathcal{E}_{\tmop{diag}}-\mathbbm{1})\rho_s
\end{eqnarray}
and
\begin{eqnarray}\label{eq:mudeph}
  \frac{\mathd}{\tmop{dt}} \rho_t= \mu( t )
 [ \mathcal{E}_{\tmop{deph}} - \mathbbm{1} ] \rho_t= \int\limits_0^t ds\, k(t-s) ( \mathcal{E}_{\tmop{deph}}-\mathbbm{1})\rho_s
\end{eqnarray}
respectively, where the quantities 
$h(t)$ and $\mu(t)$ are defined by Eqs.~\eqref{eq:h} and \eqref{eq:gamma} with 
respect to the same memory kernel $k(t)$. Note that the first evolution is always Markovian, i.e. the corresponding dynamical map is CP-divisible, as the hazard rate $h(t)$ is always positive. On the contrary, the quantity $\mu(t)$ in the latter local master equation can typically become negative, which results in
a non-Markovian dynamics. Notably, in both cases the evolution after the Redfield-like approximation is CP-divisible, since the resulting time dependent rates are positive, 
as will be shown in the next Section.

Summarising the examples considered in this Section about the open-system
dynamics of a two-level system, we can conclude what follows. If we start from either
a local or a non-local
generator, where the Lindblad operator terms describe only transitions between the levels, a term involving dephasing appears when moving to the other generator. On the contrary,
if we start from either a local or a non-local generator containing only terms which do not induce
transition among the levels, as $\mathcal{E}_{\tmop{diag}} - \mathbbm{1}$ or
$\mathcal{E}_{\tmop{deph}} - \mathbbm{1}$, no new term will appear when moving from one generator
to the other.
\\

\section{Interplay between classical memory kernel and operator contribution}\label{sec:interplay}

We will now consider two situations which put into evidence the delicate interplay between expression of the classical memory kernel and quantum operator structure. In the first instance we will consider a Redfield-like approximation which only affects the classical part and leads to a well-defined dynamics independently of the operator contribution, investigating the role of the approximation. In particular it appears that the resulting evolution is always Markovian according to either criterion, independently of the properties of the exact dynamics.  We will further put into evidence a surprising feature corresponding to a complete change of the dynamical behavior for operator contributions simply differing by a multiplying factor and used along with the very same classical kernel. Also this property can be better understood relying on the damping-basis analysis performed in Section \ref{sec:Lindblad}.

\subsection {Local approximation of semi-Markov evolution}\label{sec:Redfield}

When the memory kernel in Eq.~\eqref{eq:nz} decays much quicker than the typical time scale of the reduced system, one can approximate the exact dynamics by the time local expression given by Eq.~\eqref{eq:red}. We call this type of approximation Redfield-like \cite{megier2020interplay}. Such an approximated dynamics can have quite different properties with respect to the original one; it might even not correspond to a proper quantum evolution, loosing the positivity property \cite{Benatti2005,Whitney2008,Hartmann2020}.
In the case of quantum semi-Markov processes given by Eq.~\eqref{eq:nzSemiMark}, on the other hand, the positivity of the approximated evolution is guaranteed. More precisely the approximation
\begin{eqnarray}\label{eq:sprinkl}
  \frac{\mathd}{\tmop{dt}} \rho_t  \approx  \int^t_0 \mathd \tau k (\tau)
  (\mathcal{E} - \mathbbm{1}) \rho_t 
   =  S (t) (\mathcal{E} - \mathbbm{1}) \rho_t 
\end{eqnarray}
describes a CP divisible dynamics, since on general grounds $S (t) \geqslant 0$. This is true independently of the waiting time distribution $f(t)$, since the integral over the memory kernel can be identified with the  renewal density, also called sprinkling distribution and obeying
the renewal equation \cite{TN_libero_mab21557723,Caceres2018,Vacchini2020a}
\begin{eqnarray}
  \label{eq:4}
  S(t)=f(t)+\int\limits_0^t d\tau f(t-\tau)S(\tau).
\end{eqnarray}
Note that $S(t)$ is always positive and, despite not being a normalised probability density, provides the conditional probability density for a jump to take place at the given instant of time irrespective of previous jumps \cite{Ross2003}. Accordingly, in this situation the coarse-graining in time characterising Eq.~\eqref{eq:red} results in a kind of diffusive limit, in which the dynamics is dictated by the accumulation of many events, described by the sprinkling density. For long times, the sprinkling distribution goes to a constant value, given by the inverse mean waiting time, and a semigroup dynamics is thus recovered. Let us once more emphasise, that the above statements are true for all processes satisfying Eq.~\eqref{eq:nzSemiMark}, and therefore also for non CP-divisible ones, irrespectively of the jump map. We conclude that for this class of evolutions, the approximated dynamics is Markovian even if this was not the case for the original dynamics, as somehow naturally expected by averaging over time.

In the case of  Eq.~\eqref{eq:hdeph} the exact dynamics is divisible and therefore Markovian, and the Redfield-like approximation obtained from Eq.~\eqref{eq:sprinkl} retains this feature. Both $h(t)$ and $S(t)$ are always positive, thus admitting the natural interpretation of time-dependent rates with which the dephasing events described by 
Eq.~\eqref{eq:deph} take place. In particular we have the following general
bound
\begin{eqnarray}
  \label{eq:sh}
  S(t)\leq h(t),
\end{eqnarray}
stating that the sprinkling density always underestimates the corresponding hazard function $h(t)$, irrespectively of the underlying waiting time distribution. The inequality is saturated only in the exponential memoryless case. The bound can be obtained from monotonicity of the survival probability and the initial condition $g(0)=1$, leading to the inequalities
\begin{eqnarray}
  f(t)=\int\limits_0^t ds k(t-s)g(s) &\geq& g(t)\int\limits_0^t ds k(t-s)=S(t)g(t)\\
   f(t)=\int\limits_0^t ds k(t-s)g(s) &\leq& g(0)\int\limits_0^t ds k(t-s)=S(t)
\end{eqnarray}
and therefore upon dividing by $g(t)$ to the bounds
\begin{eqnarray}
  \label{eq:shsg}
  S(t) \leq h(t) \leq\frac {S(t)}{g(t)}.
\end{eqnarray}
Note, however, that the upper bound for long times diverges, as
typically the survival probability $g(t)$ goes to zero for long times.

\begin{center}
\minipage{0.80\textwidth}%
  \includegraphics[width=0.8\linewidth]{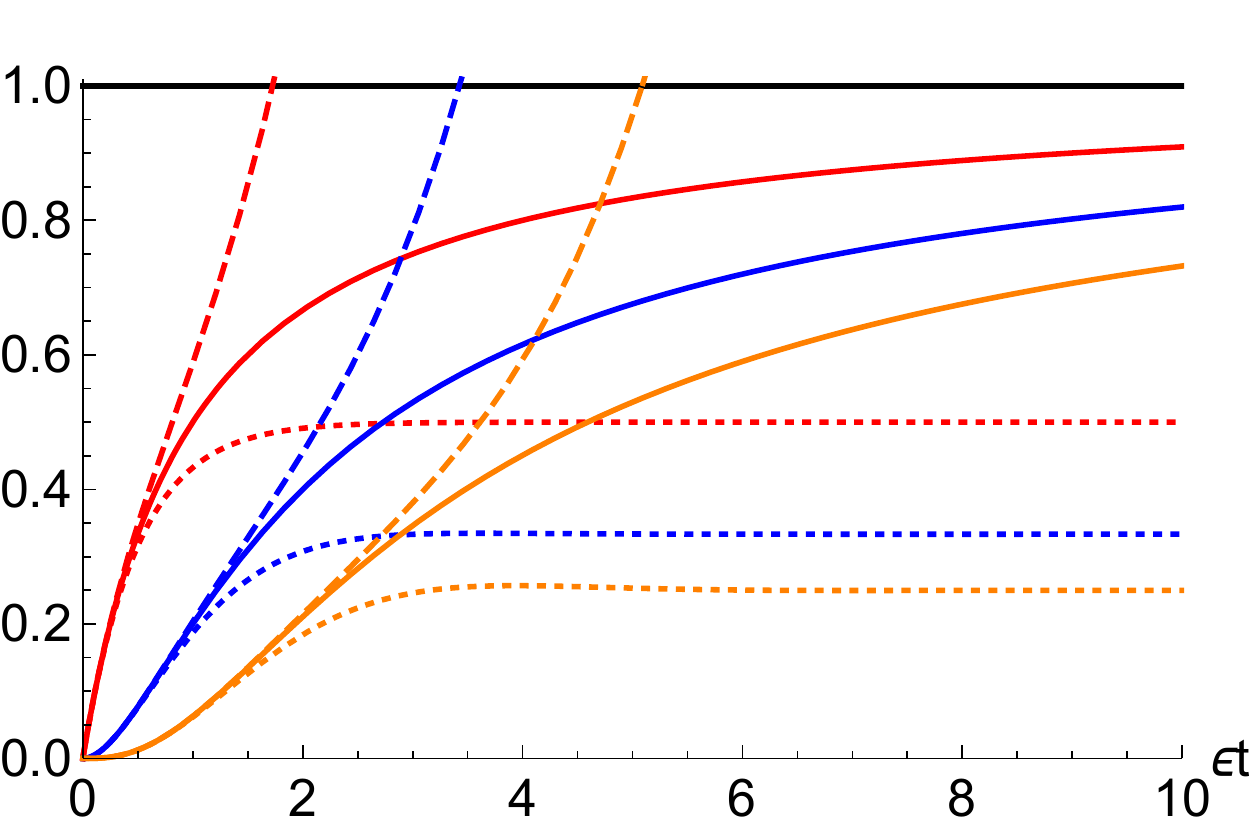}
  \endminipage
      \hspace*{-1.5cm}
  \minipage{0.17\textwidth}%
  \includegraphics[width=0.9\linewidth]{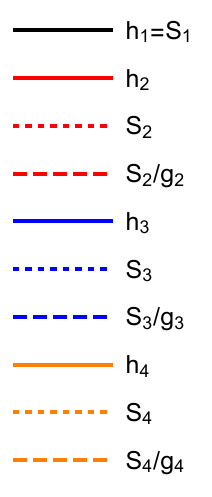}
  \endminipage
    \hspace*{0.2cm}
    \captionof{figure}{The comparison of hazard rates $h_n(t)$ (solid lines) and their lower and upper bounds, given by the sprinkling densities $S_n(t)$ (dotted lines) and the sprinkling densities over the survival probabilities $g_n(t)$ (dashed lines) respectively, for the first four Erlang probability distributions (in ascending order: black, red, blue, orange lines). For $n=1$, (black) solid and dotted lines coincide; note that in the latter case the Redfield-like approximation is exact.
    }
  \label{fig:4}
  \end{center}

To illustrate these bounds for the hazard rate $h(t)$, we have evaluated the quantities for a family of Erlang distributions $\{f_n(t)\}_{n=1,2,3,4}$, that is waiting time distribution obtained by considering the convolution of $n$ exponential distributions with the same rate parameter $\epsilon$ \cite{Ross2003}.
The $n$-th order Erlang distribution takes a simple expression in Laplace domain 
\begin{eqnarray*}
\tilde{f}_n(u)=\left(\frac{\epsilon}{\epsilon + u }\right)^n.
\end{eqnarray*}
For $n=1$, 
corresponding to the Markovian waiting time distribution of exponential type, the hazard rate is constant, $h_1(t)=\epsilon$, and the approximation is exact: $h_1(t)=S_1(t)$, see Fig. \ref{fig:4}, where
times are taken in units of  $1/\epsilon$. Asymptotically, the larger $n$ is the worse the approximation is, as following limits can be obtained
\begin{align*}
\lim\limits_{t\rightarrow \infty} h_n(t)= \epsilon, && \lim\limits_{t\rightarrow \infty} S_n(t)= \frac{\epsilon}{n},
\end{align*}
where $n/\epsilon$ is the mean waiting time or first moment of the Erlang distribution of order $n$.

\subsection{Relationship between memory kernels}\label{sec:Multiplication}

The striking difference in the behavior of the
time
evolutions described by Eqs.~{\eqref{eq:hdeph} and {\eqref{eq:mudeph}
-- only the former is always CP-divisible, but both of them are
so when approximated via the Redfield-like approximation --
is all
the more revealing of the complicated and counterintuitive interplay between
the operator contribution and the classical
functions characterising the renewal process description of the distribution of interaction events. 
In fact,
the Lindblad operator 
forms associated to the two
jump maps Eqs.~(\ref{eq:deph}) and (\ref{eq:diag}) according to
Eq.~(\ref{eq:5}) simply differ by an overall multiplying factor of two
\begin{eqnarray}\label{eq:factor2}
  \mathcal{L}_{\tmop{deph}} = 2\mathcal{L}_{\tmop{diag}} .
\end{eqnarray}

Such an apparently small discrepancy between the two generators can actually
have a major impact on the relation between the local and non-local
master equation, as well as on the possible CP-divisibility of the
corresponding dynamical map. The origin of such a difference can be found in
the different eigenvalues characterizing the corresponding damping-basis
decompositions, which shows the relevance of the analysis of Section
\ref{sec:TCLvsNZOld}. First, note that we are in the situation considered in
Eqs.~{\eqref{eq:2}} and \eqref{eq:6}}, i.e. both $\mathcal{L}_{\tmop{diag}}$ and
$\mathcal{L}_{\tmop{deph}}$ have a single non zero eigenvalue. Explicitly, they both have
a
doubly degenerate eigenvalue zero w.r.t. the eigenvectors $\mathbbm{1}$ and
$\sigma_z$, and a doubly degenerate eigenvalue, respectively $\ell _{\tmop{diag}} = - 1$ and $\ell_{\tmop{deph}} = - 2$,
w.r.t. to the eigenvectors $\sigma_x$ and $\sigma_y$.
As a consequence, the non-local equations are generated by
\begin{eqnarray}
  \mathcal{K}^{\tmop{NZ}}_{t, \tmop{diag}} & = & m^{\tmop{NZ}}_{\tmop{diag}}(t)
   (\mathcal{E}_{\tmop{diag}} - \mathbbm{1}) , \\
  \mathcal{K}^{\tmop{NZ}}_{t, \tmop{deph}} & = & m^{\tmop{NZ}}_{\tmop{deph}}(t)
   (\mathcal{E}_{\tmop{diag}} - \mathbbm{1}).  \label{eq:multipl}
\end{eqnarray}
Both $m^{\tmop{TCL}}_{\tmop{diag}} (t)$ and $m^{\tmop{TCL}}_{\tmop{deph}} (t)$
will be fixed by Eqs.(\ref{eq:mnz}), (\ref{eq:tcltcl}) and (\ref{eq:mtcl}), but, crucially, one
referred to $m^{\tmop{NZ}}_{\tmop{diag}} (t) = - k (t)$ and the other to
$m^{\tmop{NZ}}_{\tmop{deph}} (t) = - 2 k (t)$, which can result in highly
non-trivial differences between the two functions of time $m^{\tmop{TCL}}_{\tmop{diag}} (t)$ and $m^{\tmop{TCL}}_{\tmop{deph}} (t)$}. In particular, as
we see in the examples above, the local coefficient
$m^{\tmop{TCL}}_{\tmop{diag}} (t)$ is always negative, while
$m^{\tmop{TCL}}_{\tmop{deph}} (t)$ can take on positive values
which corresponds, respectively, to
CP-divisible and non CP-divisible evolutions\footnote{Note, that the corresponding decay rates $\gamma_{\tmop{diag}/\tmop{deph}}(t)$ are in these cases positive and negative, respectively, because of the negativity of the eigenvalues $\ell _{\tmop{diag}/\tmop{deph}}$.}. 
Indeed, the evolution given by Eq.~(\ref{eq:diag}) leads to a monotonic decay of coherences, i.e. the coherences  $\rho^{10/01}_t$ in the eigenbasis of $\sigma_z$ evolve as $\rho^{10/01}_t=c(t)\rho^{10/01}_0$, where $c(t)$ is monotonically decreasing function, where for the
dephasing operation described by Eq.~(\ref{eq:deph}) revivals may occur (see Fig. \ref{fig:2} for a visualisation on an example of first four Erlang probability distributions), a signature of non-Markovianity for this kind of dynamics \cite{Haase2019}.

\begin{center}
\minipage{0.80\textwidth}%
  \includegraphics[width=0.8\linewidth]{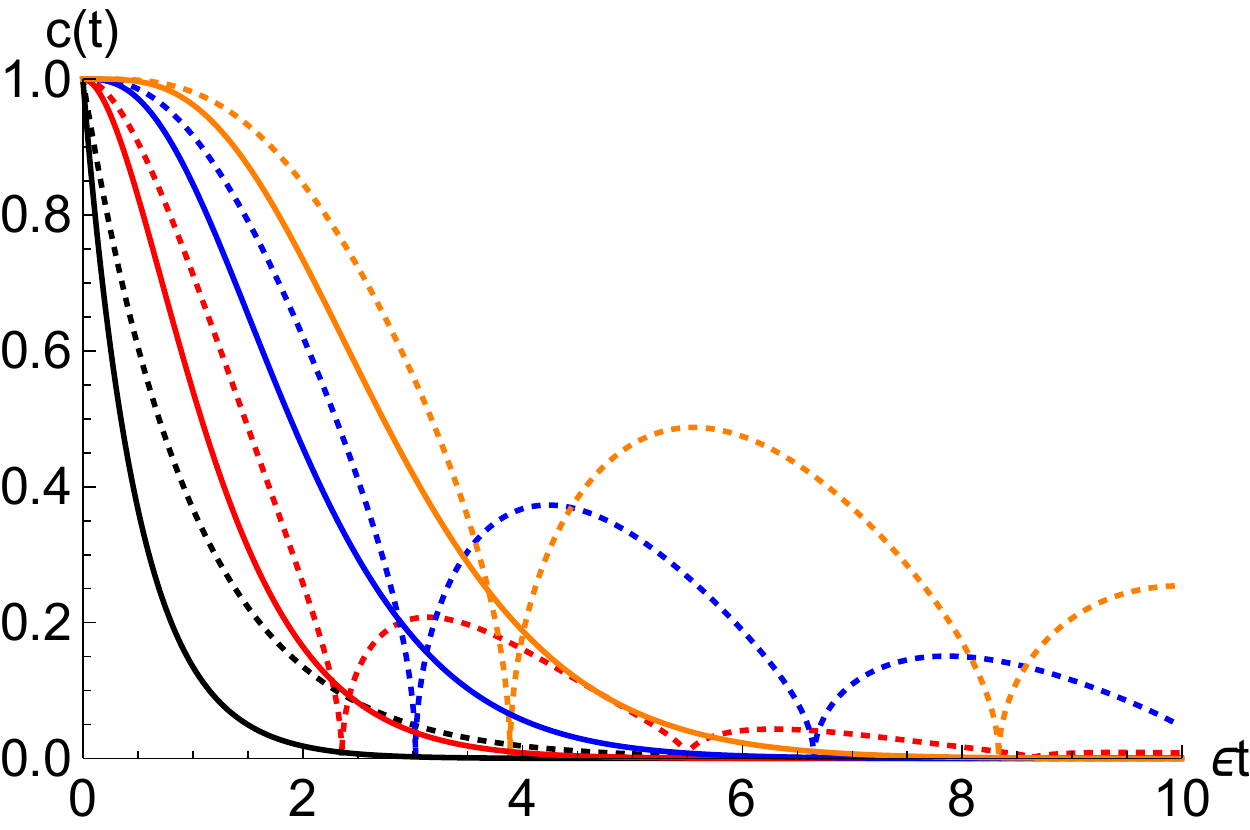}
  \endminipage
      \hspace*{-1.5cm}
  \minipage{0.17\textwidth}%
  \includegraphics[width=\linewidth]{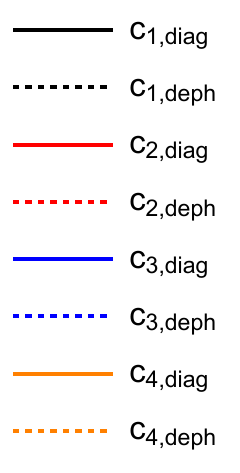}
  \endminipage
    \hspace*{0.2cm}
  \captionof{figure}{The decay of coherences ($\rho^{10/01}_t=c(t)\rho^{10/01}_0$ in the eigenbasis of $\sigma_z$) for diagonalizing (solid lines) and dephasing (dotted lines) evolutions given by, respectively, Eqs. (\ref{eq:diag}) and ~(\ref{eq:deph}), for the first four Erlang probability distributions (in ascending order: black, red, blue, orange lines).
  }
  \label{fig:2}
    \end{center}

\section{Conclusions}\label{sec:Concl}

Different types of master equations, though in principle equivalent, reveal different information about the underlying dynamics. Accordingly, the capability to obtain one of them from the other can be highly beneficial.
Here, we exploit a newly introduced link between the local and the non-local description of quantum dynamics, based on the damping-basis representation.
We focus on a class of quantum semi-Markov processes and we show that the 
different features of the two types
of master equations are strictly related to
identifying functions of the associated classical random process. 
In particular, we analyse the emergence of a dephasing term in the dynamics of an open
two-level system, when moving from
one master equation to the other. As a matter of fact, such a new term only occurs
when the corresponding waiting time distribution is classically non-Markovian, i.e. not of exponential type.
In the non-Markovian case it can also happen that the different behavior between coherences and populations can be traced back to distinct but related classical memory kernels, whose appearance is only evident in the non-local description.
We also consider the Redfield-like approximated dynamics, which is obtained via
a proper coarse graining in time. As the approximated damping rate has the meaning of renewal density, one can conclude that the Redfield-like approximation always leads to a CP-divisible evolution, as we illustrate on two, strikingly different, evolutions. The description based on the damping-basis representation further reveals the non-trivial interplay between
the operatorial structure and the classical functions fixing the considered semi-Markov evolutions.
Future studies will investigate how these classical functions can be modified in a manner that still leads to a proper quantum dynamics.

%
%
%
%

\begin{acknowledgments}
This research was funded by the UniMi Transition Grant H2020. N.M. was funded by the Alexander von Humboldt Foundation in form of a Feodor-Lynen Fellowship. A.S. and B.V. was funded by MIUR from the FFABR project. B.V. was funded by FRIAS, University of Freiburg and IAR, Nagoya University from the Joint Project ``Quantum Information Processing in Non-Markovian Quantum Complex Systems''.
\end{acknowledgments}

\bibliographystyle{unsrtnat}
\bibliography{bibliog}

\end{document}